\documentclass[12pt]{article}
\usepackage{epsfig}
\def\p2t{p_{2\bot}}
\def\q2t{q_{2\bot}}
\begin{document}
\rightline{LUNFD6/(NFFL- 7188) 2000}
\def\emline#1#2#3#4#5#6{%
       \put(#1,#2){\special{em:moveto}}%
       \put(#4,#5){\special{em:lineto}}}
\begin{center}
\vspace{1cm}
{\Large\bf
 HEAVY QUARK PHOTOPRODUCTION\\[5mm]
 IN THE SEMIHARD QCD APPROACH AND\\[5mm]
THE UNINTEGRATED GLUON DISTRIBUTION} \\[5mm]
\vspace{1cm}
{\bf    A.V. Lipatov}
\footnote{\parbox[t]{10cm}{Electronic address: artem$_-$lipatov@mail.ru}}\\
{\sl Department of Physics,\\ M.V. Lomonosov~Moscow~State~University,\\
119899~Moscow,~Russia}\\
\vspace{0.2cm}
{\bf V.A. Saleev}
\footnote{\parbox[t]{10cm}{Electronic address: saleev@info.ssu.samara.ru}}\\
{\sl Department of Physics,\\ Samara~State~University,\\
443011~Samara,~Russia}\\
\vspace{0.2cm}
{\bf N.P.~Zotov}
  \footnote{\parbox[t]{10cm}{Electronic address: zotov@theory.npi.msu.su}}\\
{\sl D.V. Skobeltsyn~Institute~of~Nuclear~Physics,\\
   M.V.Lomonosov~Moscow~State~University,\\
~119899~Moscow,~Russia}\\
  \end{center}
 \vspace*{1cm}~\\

\begin{abstract}
Processes of heavy quark photoproduction at HERA energies
are considered using the semihard
($k_{\bot}$ factorization) QCD approach with emphasis
of the BFKL dynamics of gluon distributions. We investigate
the dependences of the total cross section of  heavy quark
 photoproduction and also $p_T$, and rapidity distributions
 on different forms of the  unintegrated gluon distribution.
We present a comparision of the theoretical results with available
H1 and ZEUS experimental data for charm and beauty photoproductions.
\end{abstract}

\newpage
\noindent
\section {Introduction}
Recently, H1 and ZEUS Collaborations have reported~\cite{r1,r2}
experimental data on the total ctoss section of inelastic
open beauty photoproduction. A comparision of these results with
NLO pQCD calculations shows that ones underestimate the cross section
at HERA energies. Therefore, it would be certainly
 reasonable to try a different way.

In the present note, we focus on the so called semihard
approach~\cite{r3,r4} (SHA),
which we had applied earlier to open heavy quark~\cite{r5}
 and $J/\Psi$~\cite{r6,r7}
photoproduction.

At the HERA energies and beyond, the interaction dynamics is governed by the
properties of parton distributions in the small $x$ region. This domain is
characterized by the double inequality $s\gg\mu^2\simeq\hat s\gg\Lambda^2$,
which shows that the typical parton interaction scale $\mu$ (mass $m_c$
or $p_T$ of heavy quark) is much higher
than the QCD parameter $\Lambda$, but is much lower than the total c.m.s.
energy $\sqrt s$. The situation is therefore classified as ``semihard''.

The resummation \cite{r3,r4,r8,r9} of the terms
$[\mbox{ln}(\mu^2/\Lambda^2)\,\alpha_s]^n$,
$[\mbox{ln}(\mu^2/\Lambda^2)\,\mbox{ln}(1/x)\,\alpha_s]^n$ and
$[\mbox{ln}(1/x)\,\alpha_s]^n$ in SHA results in the unintegrated parton
distributions $\Phi_i(x,q_{T}^2,\mu)$, which determine the probability
to find a parton of type $i$ carrying the longitudinal momentum fraction $x$
and transverse momentum $q_T$ at the probing scale $\mu^2$.
They obey the BFKL
equation~\cite{r10} and reduce to the conventional parton densities $F_i(x,\mu^:)$
once the $q_{T}$ dependence is integrated out:
\begin{equation} \label{kt}
\int_0^{\mu^2}\!\!\Phi_i(x,q_{T}^2,\mu^2,\ Q^2_0)\;dq_{T}^2=
x\,F_i(x,\mu^2,\ Q^2_0).
\end{equation}

To calculate the cross section of a physical process, the unintegrated
functions $\Phi_i$  have to be convoluted
with off-mass shell matrix elements
corresponding to the relevant partonic subprocesses.
In the off-mass shell matrix element the virtual gluon polarization
tensor is taken in the form of the SHA prescription \cite{r3,r4}:
\begin{equation}
L^{(g)}_{\mu\nu}=\overline{\epsilon_2^{\mu}\epsilon_2^{*\nu}}
  =p^\mu p^\nu x^2/|q_T|^2
  =q_T^\mu q_T^\nu/|q_T|^2
\end{equation}

  In ref.~\cite{r5} was used
phenominological
parametrization for the unintegrated gluon distribution including an
arbitrary normalization constant ($K-$ factor), which was obtained
from a fit to $b\bar b -$pair production at the Tevatron \cite{r4}.
In our recent paper~\cite{r7} we investigate the sensitivity
of heavy quarkonium photoproduction to
different gluon distributions. Special attention was given to
the unintegrated gluon distributions obtained from BFKL evolution
equation. In this paper we study the sensitivity of the total
cross section of inelastic charm abd beauty photoproduction
to these  unintegrated gluon distributions.
 The outline of our paper is the following: In sect. 2, we give
the formulas for the cross sections of heavy quark photoproduction
in the SHA of QCD. Then, in sect. 3, we describe the unintegrated
distributions which we use for our calculations.
In sect. 4, we present
the results of our calculations. Finally, in sect. 5, we give some
conclusions.

\section{SHA QCD cross section for heavy quark photoproduction}

We calculate the total and differential cross sections (the
$p_{\bot}$ and rapidity distributions) of charm and beauty quark photoproduction
via the photon-gluon fusion QCD subprocess (Fig.1) in the framework
of the SHA.

First of all we take into account the transverse momentum of gluon
$\vec{q}_{2\bot}$, its the virtuality $q^2_{2}=-\vec{q}^2_{2\bot}$ and
the alignment of its polarization vectors along its transverse momentum
such as $\epsilon_{\mu} = q_{2\bot\mu}/\mid\vec{q}_{2\bot}\mid$
~\cite{r3,r4,r8,r9}.
Let us define Sudakov variables of the process $ep\to Q\bar Q X$
(Fig.1):
\begin{eqnarray}
p_1&=&\alpha_1 P_1+\beta_1 P_2+p_{1\bot}\qquad
p_2=\alpha_2 P_1+\beta_2 P_2+\p2t\nonumber\\
q_1&=&x_1P_1+q_{1\bot}\qquad
q_2=x_2P_2+\q2t
\end{eqnarray}
where
$$p_1^2=p_2^2=M^2,\qquad q_1^2=q_{1\bot}^2,\qquad q_2^2=\q2t^2,$$
$p_1$ and $p_2$ are 4-momenta of the heavy quarks, $q_1$ is 4-momentum
 of the photon, $q_2$ is 4-momentum of the gluon, $p_{1\bot},~~p_{2\bot},~~
 q_{1\bot},~~q_{2\bot}$ are transverse 4-momenta of these ones.
 In the center of mass frame of colliding particles we can write
 $P_1=(E,0,0,E)$, $P_2=(E,0,0,-E)$, where $E=\sqrt s/2$, $P_1^2=P_2^2=0$
  and $(P_1P_2)=s/2$.
Sudakov variables are expressed  as follows:
\begin{eqnarray}
\alpha_1 &=&\frac{M_{1\bot}}{\sqrt s}\exp(y_1^\ast)\qquad
\alpha_2 =\frac{M_{2\bot}}{\sqrt s}\exp(y_2^\ast)\nonumber\\
\beta_1 &=&\frac{M_{1\bot}}{\sqrt s}\exp(-y_1^\ast)\qquad
\beta_2 =\frac{M_{2\bot}}{\sqrt s}\exp(-y_2^\ast),
\end{eqnarray}
where $M_{1,2\bot}^2=M^2+p_{1,2\bot}^2$, $y_{1,2}^{\ast}$ are rapidities
of heavy quarks, M is heavy quark mass.

From conservation laws we can easly obtain the following conditions:
\begin{equation}
q_{1\bot}+\q2t=p_{1\bot}+\p2t,\qquad
 x_1=\alpha_1 +\alpha_2,\qquad
 x_2=\beta_1 +\beta_2
\end{equation}
The differential cross section of heavy quark  photoproduction has the
following form
\begin{equation}
\frac{d\sigma}{d^2p_{1\bot}}(\gamma p\to Q\bar Q X)=
\int dy_1^{\ast}\frac{d^2q_{2\bot}}{\pi}
\frac{\Phi_p(x_2,q_{2\bot}^2)|\bar M|^2}
 {16\pi^2(sx_2)^2\alpha_2}
\end{equation}

The matrix element $\bar M$ for a subprocess $\gamma g^{\ast} \to
q\bar q$ depends on the virtuality of the gluon and differs from the one
of the usual parton model. For the square of this matrix element we used the
following form~\cite{r8}:
\begin{eqnarray}
|\bar M|^2=16\pi^2e_Q^2\alpha_s\alpha_{em}(x_2s)^2[\frac{\alpha_1^2+\alpha_2^2}
 {(\hat t-M^2)(\hat u-M^2)}-\frac{2M^2}{q_{T}^2}
(\frac{\alpha_1}{\hat u-M^2}-
 \frac{\alpha_2}{\hat t-M^2})^2],
\end{eqnarray}
where $\hat s,~\hat t,~\hat u$ are usual
Mandelstam variables of partonic subprocess $\gamma g^{\ast} \to q\bar q$.

\section{Unintegrated gluon distribution}

In this paper we used the different parametrizations for the
unintegrated gluon distribution. First, as in the publication
~\cite{r5}, we used the following phenomenological parametrization
(LRSS-parametrization)~\cite{r3,r4}:
\begin{equation}
\Phi (x,\vec{q}_T\,^2)=\Phi_0\frac{0.05}{0.05 + x}(1-x)^3f_1(x,\vec{q}_T\,^2),
\end{equation}
where
\begin{equation}
f_1(x,\vec{q}_T\,^2)=\cases{
                         1,& $\mbox{if} \quad\vec{q}_T\,^2 \le q_0^2(x)$,\cr
(q_0^2(x)/\vec{q}_T\,^2)^2,& $\mbox{if} \quad\vec{q}_T\,^2 > q_0^2(x)$}
\end{equation}
with $q_0^2(x)=q_0^2+\Lambda^2\exp(3.56\sqrt{ln(x_0/x)})$,
$q_0^2=2GeV^2$, $\Lambda=56 MeV$, $x_0=1/3$. The  value  of the
parameter $q_0^2(x)$ can be considered as a new typical transverse
momentum  of the
partons in the parton cascade which leads to natural
infrared cut-off in semihard approach. The normalization  factor
$\Phi_0$ of the  structure  function  $\Phi(x,\vec q_T\,^2)$   was
obtained  in ~\cite{r4}, where beauty production at CDF energy was
described, $\Phi_0 = 0.97$ mb.

 The effective gluon distribution $xG(x,\mu ^2)$, which was obtained
from eq. (7)-(8), and eq. (1) increases at not very low $x (0.01< x< 0.15)$ as
\begin{equation}
 xG(x,\mu ^2)\sim x^{-\Delta},
 \end{equation}
  where $\Delta \approx 0.5$ corresponds to the QCD  pomeron singularity
  given by summation of leading logarithmic contributions
($\alpha_s ln\frac{1}{x})^n$~\cite{r10}. This increase continues up to
$x=x_0$,
    where $x_0$ is a solution of the equation $q_0^2(x_0)=\mu ^2$.
    In the region $x<x_0$, there is saturation of the gluon
distribution
  function: $xG(x,\mu ^2)\approx \Phi_0\mu ^2$.

A second parametrization is based on the numerical solution of the BFKL
evolution equations~\cite{r11} (RS--parametrization).
The solution has the following form~\cite{r11}:
\begin{eqnarray}
\Phi(x, q^2) = \frac{a_1}{a_2 + a_3 +a_4}[a_2 + a_3 (\frac{Q_0^2}{q^2})
+ (\frac{Q_0^2}{q^2})^2 +
\nonumber \\
 \alpha x +
 \frac{\beta}{\epsilon + \ln (1/x)}]
C_q [\frac{a_5}{a_5 + x}]^{1/2}[1 -
\nonumber \\
 a_6 x^{a_7}\ln{(q^2/a_8)}]
(1 + a_{11} x) (1  - x)^{a_9 + a_{10}\ln (q^2/a_8)},
\end{eqnarray}
where
\begin{equation}
C_q = \cases{
1,          &$\mbox{if}\; q^2  < q_0(x)$, \cr
q_0(x)/q^2, &$\mbox{if}\; q^2  > q_0(x)$.}
\end{equation}
 All parameters (see~\cite{r11}) $(a_1 - a_{11}, \alpha, \beta$ and
$\epsilon)$ were found by minimization of the differences between
left hand and right-hand of the BFKL-type equation for
unintegrated gluon distribution $\Phi (x, q^2)$, $Q_0^2 =$4 GeV$^2$.

Finally we also use the results of a BFKL-like
 parameterization \footnote{ Of course LRSS  and RS parametrizations
are BFKL - type too.} of
the unintegrated gluon distribution
$\Phi(x,q^2_{T},\mu^2)$, according to the prescription given in~\cite{r12}.
The proposed method lies upon a straightforward perturbative solution of
the
BFKL equation where the collinear gluon density $x\,G(x,\mu^2)$
from the standard GRV set~\cite{r12} is used as
the boundary condition in the integral form (1).
Technically, the unintegrated gluon density is calculated as a convolution
of collinear gluon density $G(x,\mu ^2)$
with universal weight factors~\cite{r12}:
\begin{equation}
 \Phi(x,q_T^2,\mu^2) = \int_x^1
 {\cal G}(\eta,q_T^2,\mu^2)\,
 \frac{x}{\eta}\,G(\frac{x}{\eta},\mu^2)\,d\eta,
\end{equation}
where
\begin{equation}
 {\cal G}(\eta,q_T^2,\mu^2)=\frac{\bar{\alpha}_s}{\eta q_T^2}\,
 J_0(2\sqrt{\bar{\alpha}_s\ln(1/\eta)\ln(\mu^2/q_T^2)}),
 \qquad q_T^2<\mu^2,
\end{equation}

\begin{equation}
 {\cal G}(\eta,q_T^2,\mu^2)=\frac{\bar{\alpha}_s}{\eta q_T^2}\,
 I_0(2\sqrt{\bar{\alpha}_s\ln(1/\eta)\ln(q_T^2/\mu^2)}),
 \qquad q_T^2>\mu^2,
\end{equation}
where $J_0$ and $I_0$ stand for Bessel functions (of real and imaginary
arguments, respectively), and $\bar{\alpha}_s=3{\alpha}_s/\pi$.
The  parameter $\bar{\alpha}_s$ is connected with
 the Pomeron trajectory intercept:
$\Delta=\bar{\alpha}_s4\ln{2}$ in the LO and
$\Delta=\bar{\alpha}_s4\ln{2}-N\bar{\alpha}_s^2$ in the NLO approximations,
respectively, where $N \sim 18$  \cite{r14}. The latter value
of $\Delta$ have dramatic
consequences for high energy phenomenology. In particular it
leads to negative values for physical cross sections~\cite{r15}.
However some resummation procedures proposed in the last years
lead to positive value of $\Delta (\sim 0.2 - 0.3)$~\cite{r16,r17}.
Therefore in our calculations with (13) we used only
the solution of LO BFKL equation and
considered $\Delta$ as free parameter varying it from 0.166 to 0.53.
\par
The presence of the two different  parameters, $\mu^2$ and $q^2_{T}$, in
eq.(12)
for  unintegrated gluon distribution
$\Phi(x,q^2_{T},\mu^2)$ refers  to the fact that
the evolution of parton densities  proceeds in two steps.
First  the DGLAP scheme~\cite{r18} is applied to evolve the
structure function from $Q^2_0$ to $\mu^2$ within the
collinear approximation.
After that  eqs. (12)-(14) are used to develop the
parton transverse momenta $q_T^2$ in correspondence with BFKL
evolution~\cite{r10}.

 This approach was used for the description of $p_T$ spectrum
of $D^*$ meson electroproduction at HERA~\cite{r19} and $J/\Psi$
photoproduction~\cite{r7}, where in the first case for
Pomeron intercept parameter was obtained the value $\Delta =
0.35$. However for the total cross section of inelastic $J/\Psi$
photoproduction $\Delta = 0.53$ is more preferable.

\section{Results of calculations}

The calculations of the heavy quark photoproduction cross section
in the SHA have been made accoding to eqns. (6) and (7) for  $\vec q_T\,^2 >
q_0^2$ GeV$^2$
and for $\vec q_T\,^2 \leq q_0^2$ GeV$^2$ we set $|\vec q_T| = $ 0 in the
matrix element of process and use one of the standard partom model (SPM).
 The choice of the critical value of parameter
$\vec q_T\,^2 = q_0^2 = 1 - 2 $ GeV$^2$ is determined by the requirement
of the small value of $\alpha _s(\vec q_T\,^2)$ in the region
$\vec q_T\,^2 > 1 - 2$ GeV$^2$, where in fact
 $\alpha _s(\vec q_T\,^2) \leq 0.26$
\footnote{The substitution $q_0^2 =1$ GeV$^2$ instead 2 GeV$^2$
essentially influences on the dependence
of cross sections from the BFKL parameterization of unintegrated
gluon distribution.}.

The results of our calculations for the cross sections of
$c \bar c$ and $b \bar b$  photoproduction  processes
 are shown in Figs. 2 - 10. Fig. 2 shows the total cross section of the
 inelastic
 $c\bar c$ photoproduction at HERA and fixed target energies as a function
 of $\sqrt s$.
 The curves 3, 4 and 5 correspond to the SHA calculations
with the RS, LRSS and BFKL unintegrated gluon distributions at
the Pomeron intercept $\Delta = 0.35$, $m_c = 1.5$ GeV and at values of
$q_0^2 = 4,\, 2$ and 1 GeV$^2$ accordingly. One can see that the SHA
curves 4 and 5 describe the data better than the SPM ones
(1 and 2). The RS
parameterization of the unintegrated gluon distribution gives the
total cross section more close to SPM result (obtained with GRV gluon
density) because of very large value of $q_0^2 = 4$ GeV$^2$.

The LRSS parameterization of the unintegrated gluon distribution
gives the cross section which stops to increase  at energy $\sqrt s
\geq 150$ GeV because of saturation effects
 accounted for by the parameterizations (8) and (9). We would like
to stress that account for saturation effects
in the region $x \leq 10^{-3}$~\cite{r5,r20}, does not contradict
the H1 and ZEUS experimental data~\cite{r21,r22}.

The BFKL parameterization (curve 5) describes the H1 and ZEUS data
very well at $m_c = 1.3$ GeV (Fig. 3)
 The curves 3, 4 and 5
obtained in the SHA show more rapid growth with energy than
those obtained in the SPM with GRV~\cite{r13}(curve 2) and MT~\cite{r23}
(curve 1) parameterizations.

In Fig.4 we show the dependence of the total cross section from
$\Delta$ parameter for the BFKL parametrization (13) - (15) at
 $m_c = 1.3$ GeV. The dependence is very weak
in contrast with the dependence of the total inelastic $J/\Psi$
photoproduction cross section \cite{r7}.

The results of our calculations for the $p_T$ and rapidity $y$ ($y = y_1^*$)
 distributions of the c quarks
 at $\sqrt s = 200 $ GeV are shown in Figs. 5 and 6.
The SHA curves 4 and 5 in Fig. 5 for the  $p_T$  distribution
are  higher than those of the SPM. It is effect of
the SHA broadening of $p_T$ spectra due to extra transverse momentum
of the initial gluons~\cite{r4}-~\cite{r6}. The SHA rapidity distributions
(Fig. 6) have  more similar behaviour in wide rapidity region.

Fig. 7 shows the total cross section of inelastic
$b\bar b$ photoproduction at HERA energies as a function
of $\sqrt s$. The curves 3, 4 and 5 correspond to the SHA calculations
with the RS, LRSS and BFKL unintegrated gluon distributions at
the Pomeron intercept $\Delta = 0.35$, $m_b = 4.75$ GeV and at values of
$q_0^2 = 4,\, 2$ and 1 GeV$^2$ accordingly. The H1~\cite{r1} and
EMC~\cite{r24} experimental data
 are described by the LRSS parameterization (curve 4) and
the BFKL parameterization (curve 5 in Fig. 8)  but
at small value of $m_b = 4.25$ GeV only for the last  parameterization.

The results of the SHA calculations for the $p_T$ and rapidity
 distributions of the b quarks
 at $\sqrt s = 200 $ GeV are shown in Figs. 9 and 10.
The differences between SHA curves and SPM ones are more pronounced
than in the c-quark case.

\section{Conclusions}

We considered the process of
inelastic heavy quark photoproduction at HERA in  the framework of the
semihard QCD approach with emphasis on
the BFKL dynamics of gluon distributions.
We investigated total cross section of inelastic
$c-$ and $b-$quark photoproduction and also $p_T$ and rapidity spectra
as a function of different
unintegrated gluon distributions. It is shown that
the description of the $c-$quark inelastic
cross section at HERA energies is achieved in the cases
 of the LRSS and BFKL parameterizations
 at $\Delta =0.35$ \footnote{ Close
 values for the parameter $\Delta$ were obtained rather in
 very different papers (see, for example, \cite{r25}
 -\cite{r28}) and in the L3 experiment \cite{r29}.}
 and at the values of $m_c = 1.5$ and 1.3 GeV
 and at $q_0^2 =2$ and 1 GeV$^2$ accordingly.

The cross section of inelastic $b-$quark photoproduction at HERA
is described by the LRSS parameterization at the value of $m_b = 4.75$ GeV
and  $\Delta = 0.35$ very well. The BFKL paremeterization with $\Delta =
0.35$ describes the H1 and EMC experimental data at $m_b = 4.25$ GeV
only.

The dependence of the total cross section of inelastic open
$c-$ and $b-$quark photoproduction on $\Delta$ parameter for the BFKL
parameterization (13) - (15) is very weak in contrast with the dependence
of the total inelastic $J/\Psi$ photoproduction cross section.

\section{Acknowledgments}

 One of us (N.Z.) would like to thank S. Baranov and
 H. Jung for numerous discussions of different aspects of the QCD SHA,
  the Elementary Particle Department of Institute of Physics
of Lund University and specially Leif J\"onsson for their warm
hospitality and the Theoretical
Department of Institute of Physics of Lund University for very useful
discussions. This work has been supported in part by the Royal Swedish
Academy of Sciences.

\newpage
 {\bf Figure captions}\\

\noindent {\bf Fig.~1~}
QCD diagram for the open heavy quark electroproduction.\\

\noindent {\bf Fig.~2~}
The total cross section of inelastic  $c \bar c$ photoproduction
as a function $\sqrt s$ \\
at $m_c = 1.5$ GeV. \\

\noindent {\bf Fig.~3~}
The total cross section of inelastic $c \bar c$ photoproduction
as a function $\sqrt s$ \\
at $m_c = 1.3$ GeV. \\

\noindent {\bf Fig.~4~}
The total cross section of inelastic  $c \bar c$ photoproduction
as a function $\sqrt s$ \\
at  $m_c = 1.3$ GeV and different values
of $\Delta$: {\bf 1} -  $\Delta = 0.166, \, {\bf 2} -  \Delta = 0.35, \\
{\bf 3} -  \Delta = 0.53$.\\

\noindent {\bf Fig.~5~}
The $p_T^2$ distribution of inelastic  $c \bar c$ photoproduction at
$\sqrt s =$200 GeV.\\

\noindent {\bf Fig.~6~}
The rapidity distribution of inelastic  $c \bar c$ photoproduction
at $\sqrt s =$200 GeV. \\

\noindent {\bf Fig.~7~}
The total cross section of inelastic  $b \bar b$ photoproduction
as a function $\sqrt s$ \\
at $m_b = 4.75$ GeV. \\

\noindent {\bf Fig.~8~}
The total cross section of inelastic  $b \bar b$ photoproduction
as a function $\sqrt s$ \\
at $m_b = 4.25$ GeV. \\

\noindent {\bf Fig.~9~}
The $p_T^2$ distribution of inelastic  $b \bar b$ photoproduction
at $\sqrt s =$200 GeV.\\

\noindent {\bf Fig.~10~}
The rapidity distribution of inelastic  $b \bar b$ photoproduction
$\sqrt s =$200 GeV.

\newpage
\begin{figure}
\begin{center}
\epsfig{figure=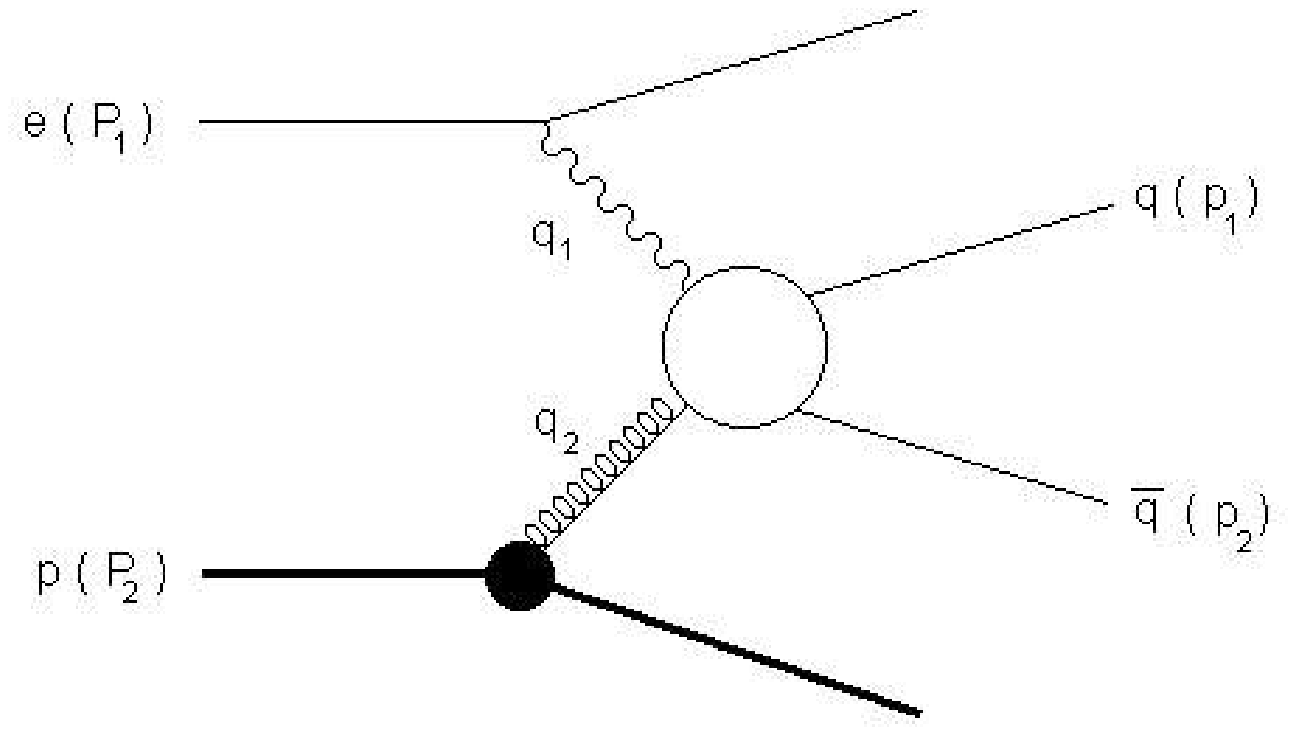,width=12cm,height=10cm}
\end{center}
\centerline{Fig.~1.}
\label{fig1}
\end{figure}
%
\begin{figure}
\begin{center}
\epsfig{figure=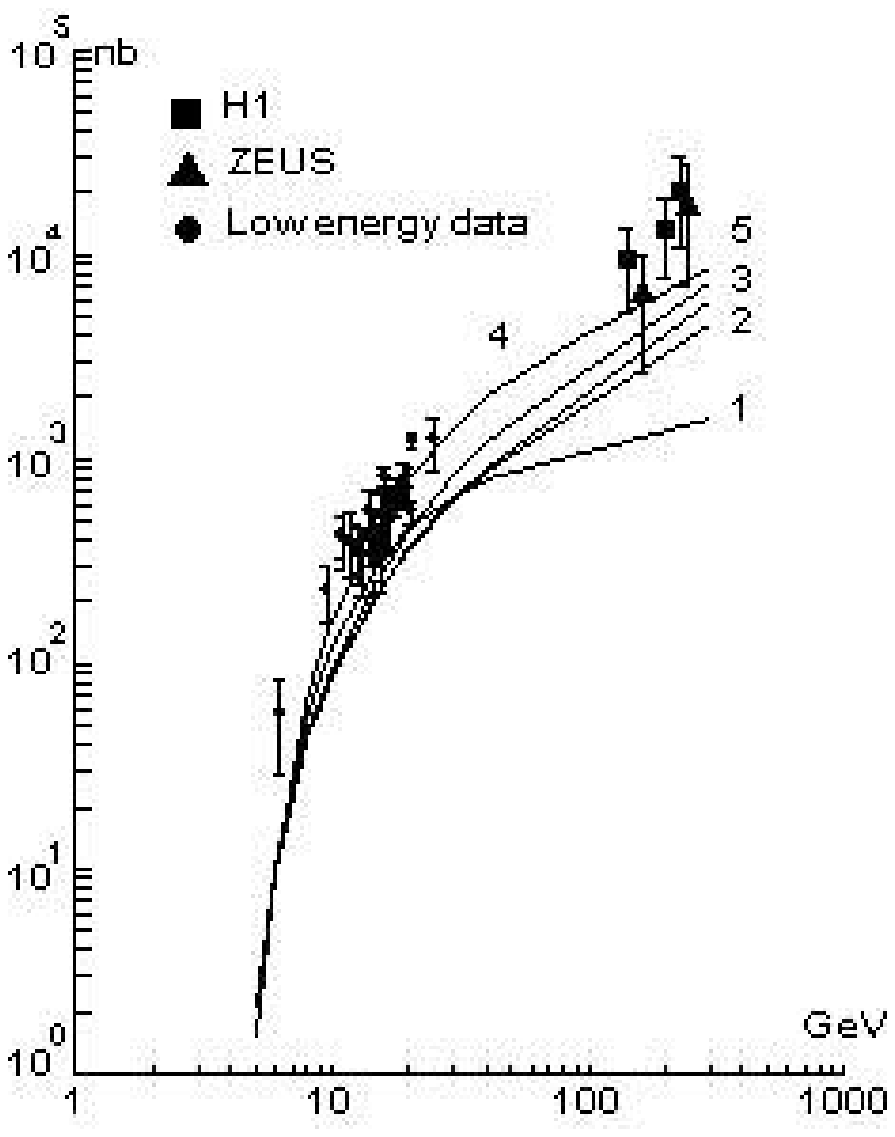,width=10cm,height=10cm}
\end{center}
\centerline{Fig.~2.}
\label{fig2}
\end{figure}
\begin{figure}
\begin{center}
\epsfig{figure=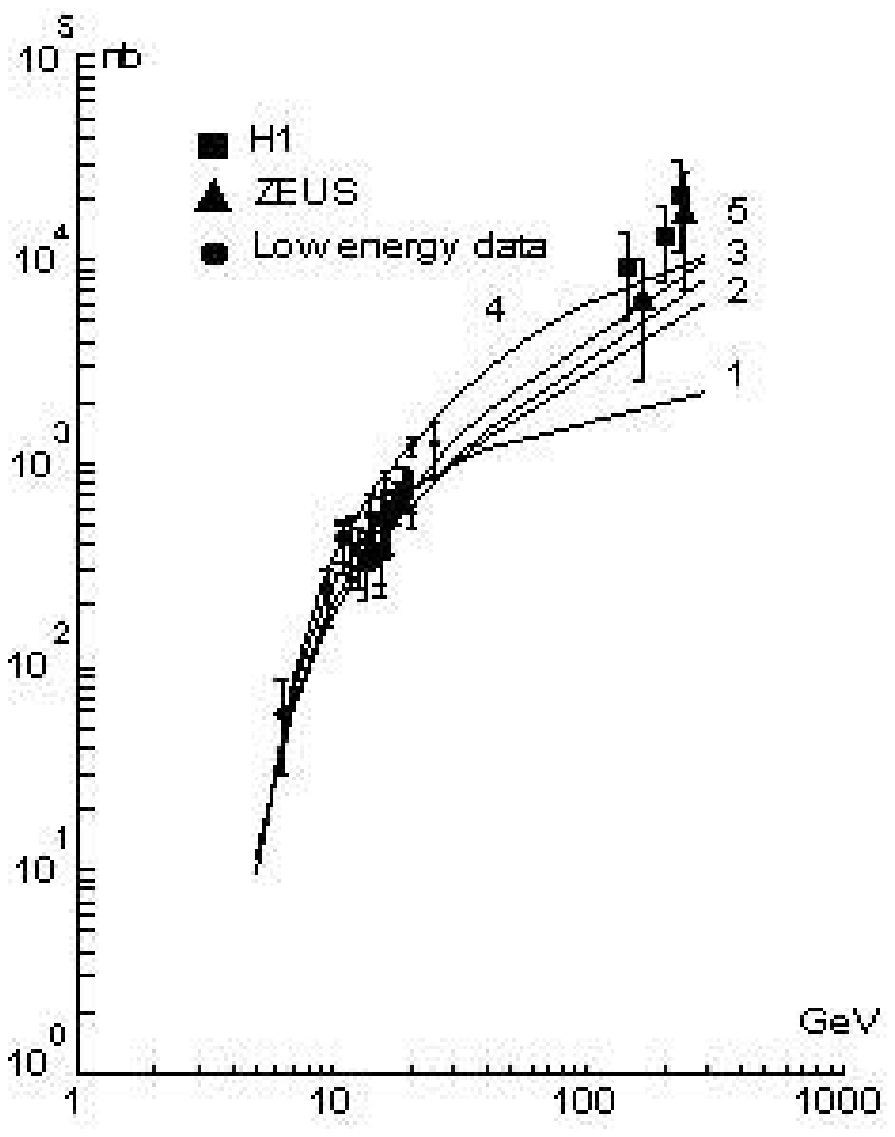,width=10cm,height=10cm}
\end{center}
\centerline{Fig.~3.}
\label{fig3}
\end{figure}
\begin{figure}
\begin{center}
\epsfig{figure=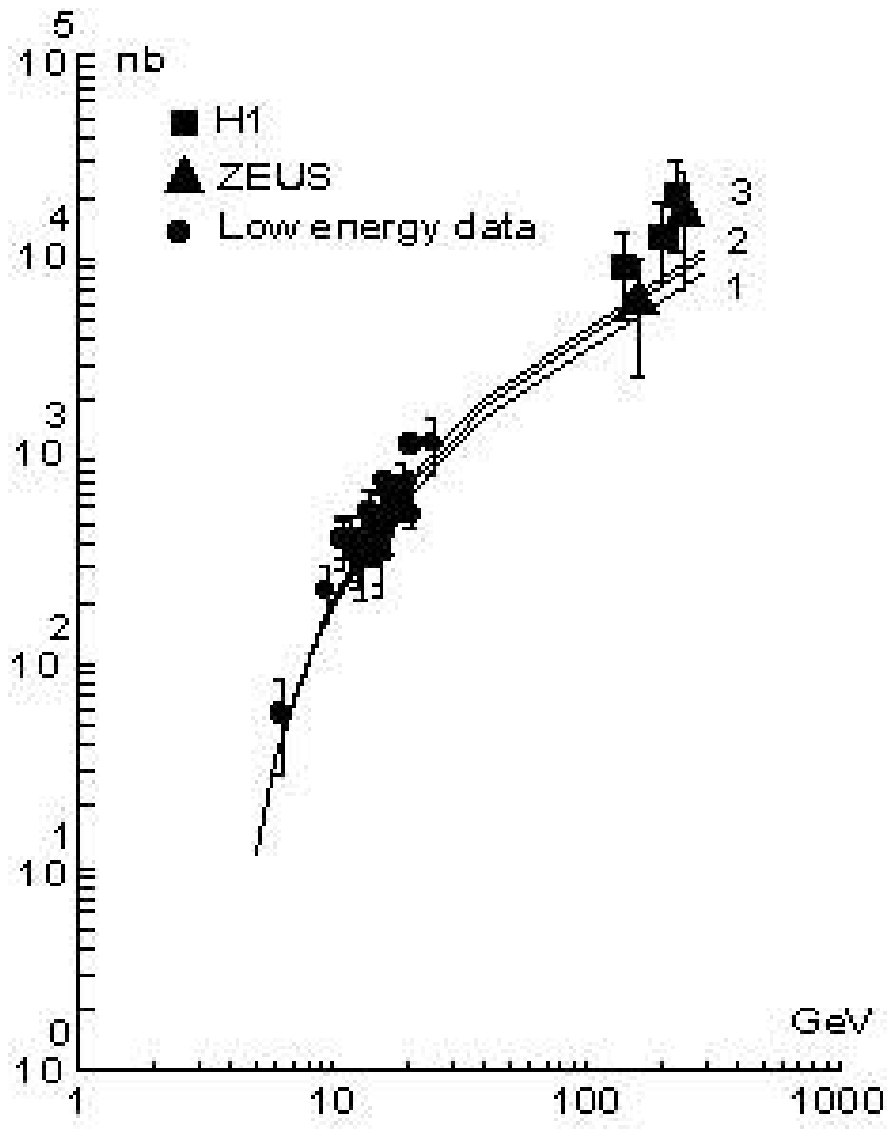,width=10cm,height=10cm}
\end{center}
\centerline{Fig.~4.}
\label{fig4}
\end{figure}
\begin{figure}
\begin{center}
\epsfig{figure=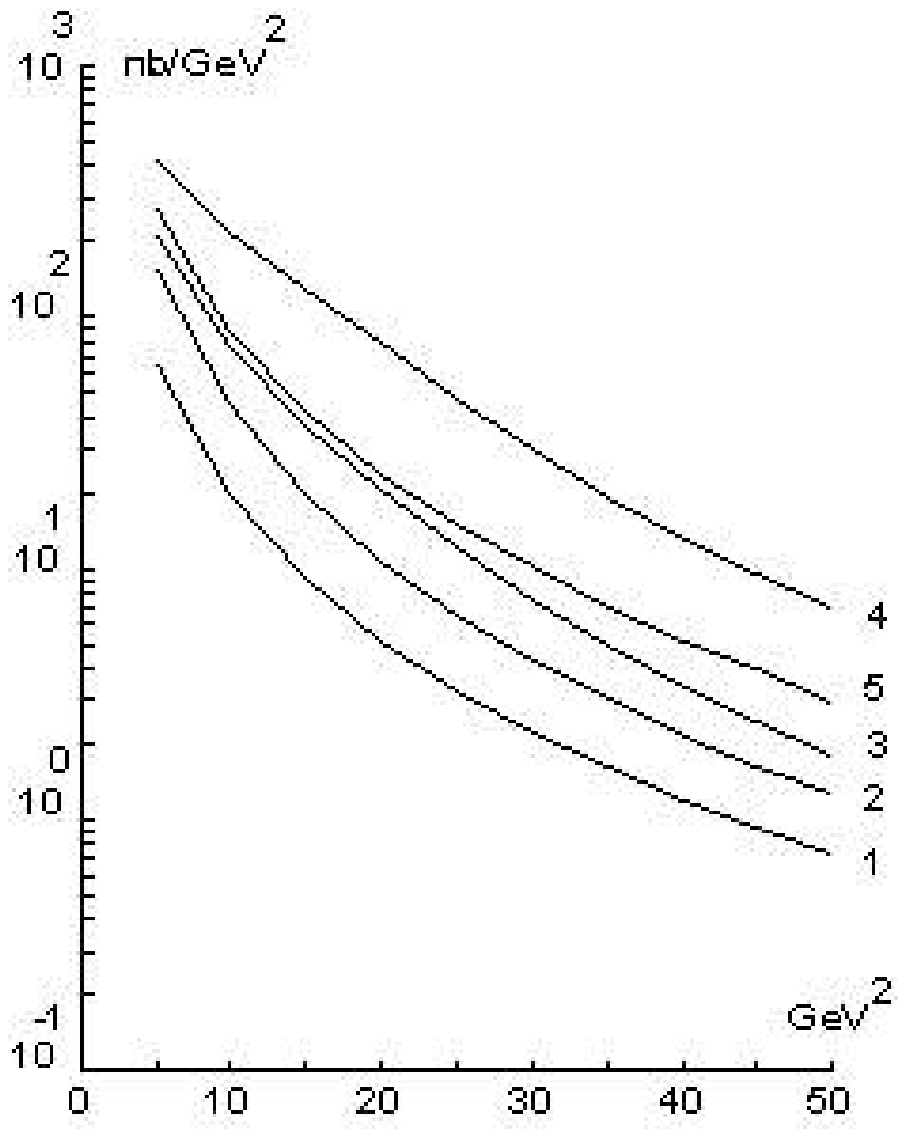,width=10cm,height=10cm}
\end{center}
\centerline{Fig.~5.}
\label{fig5}
\end{figure}
\begin{figure}
\begin{center}
\epsfig{figure=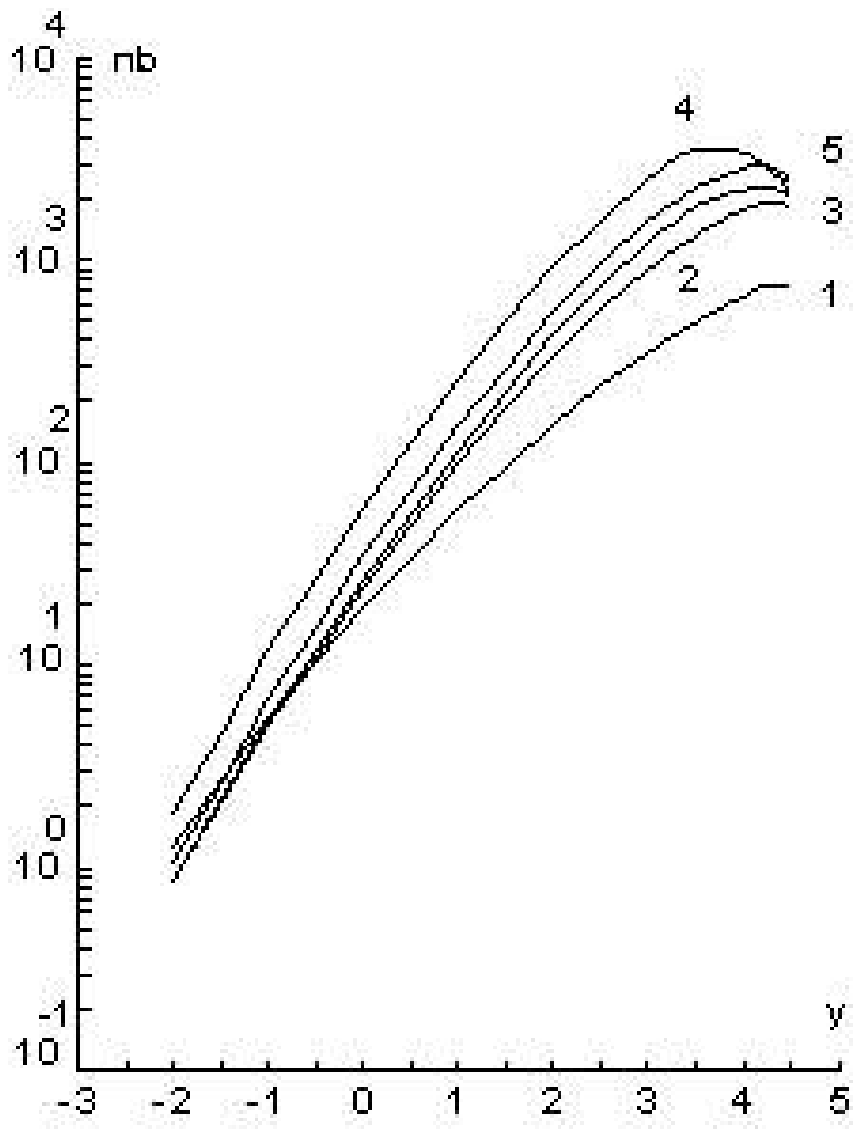,width=10cm,height=10cm}
\end{center}
\centerline{Fig.~6.}
\label{fig6}
\end{figure}
\begin{figure}
\begin{center}
\epsfig{figure=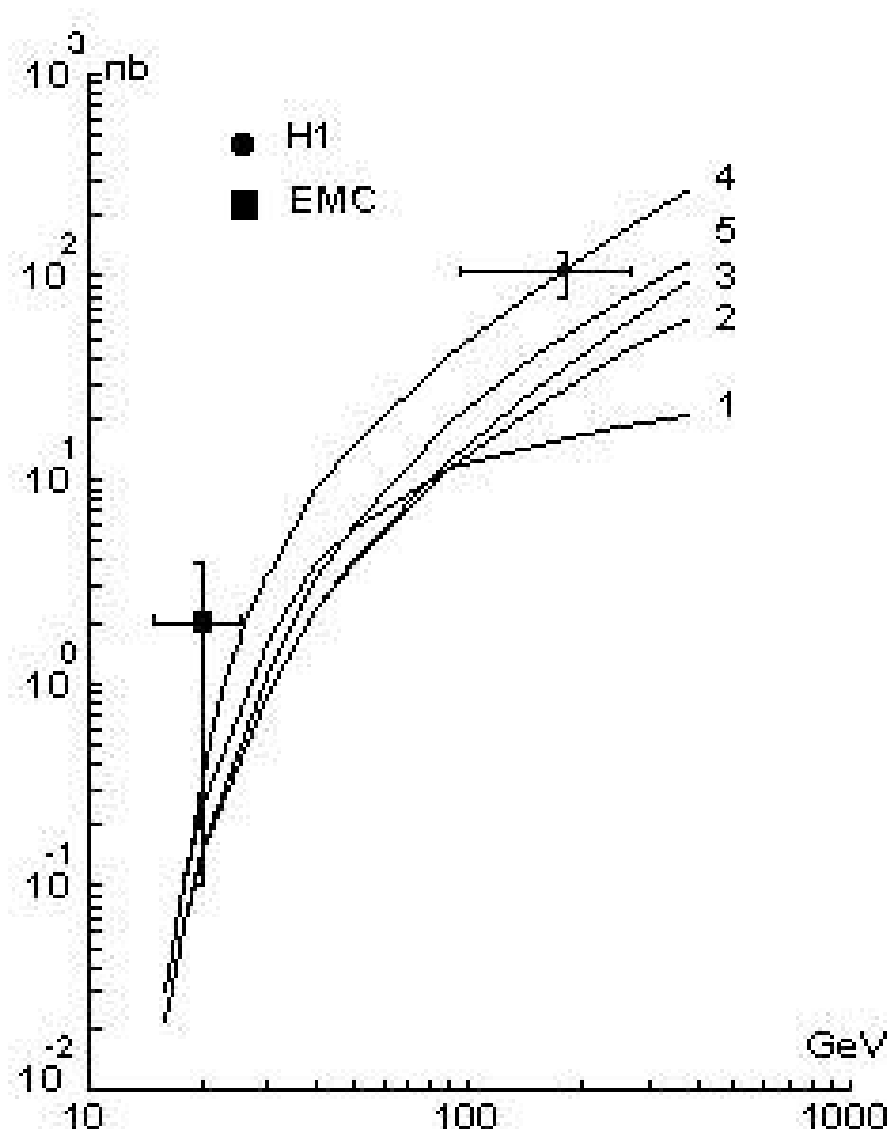,width=10cm,height=10cm}
\end{center}
\centerline{Fig.~7.}
\label{fig7}
\end{figure}
\begin{figure}
\begin{center}
\epsfig{figure=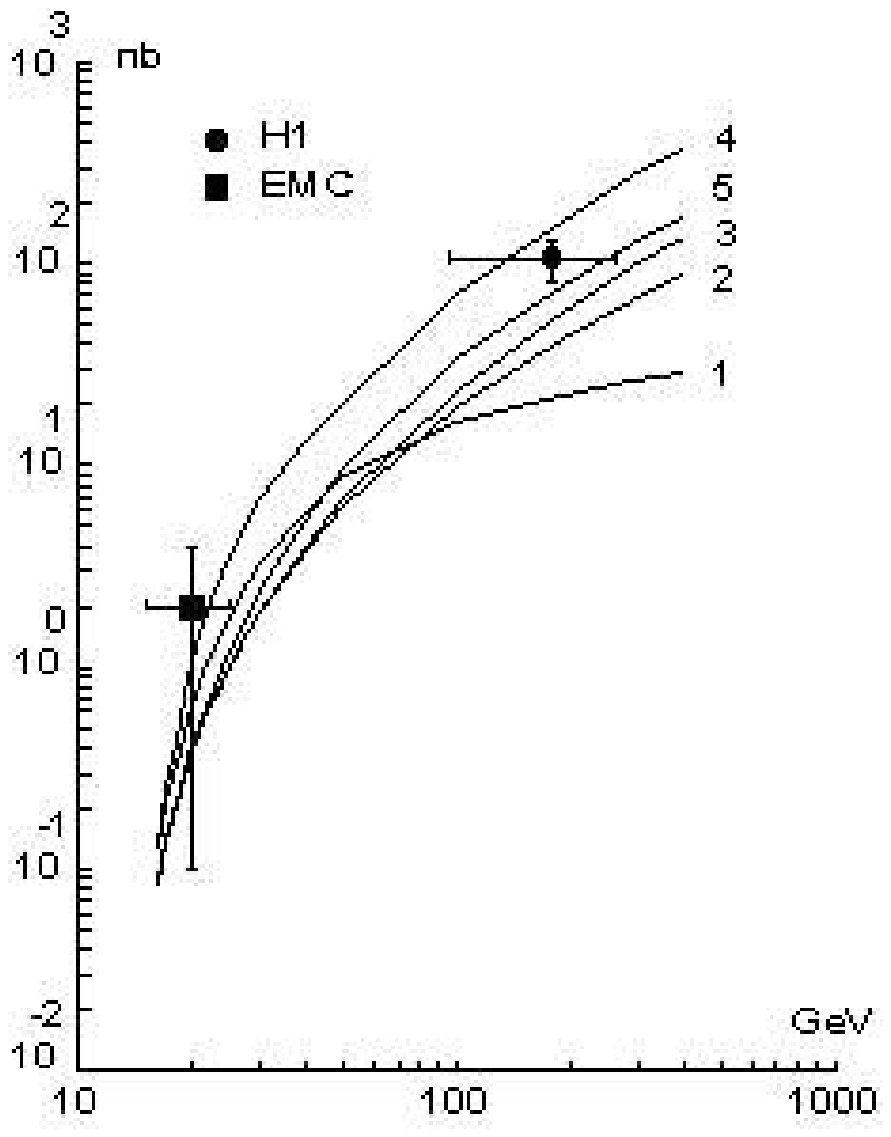,width=10cm,height=10cm}
\end{center}
\centerline{Fig.~8.}
\label{fig8}
\end{figure}
\begin{figure}
\begin{center}
\epsfig{figure=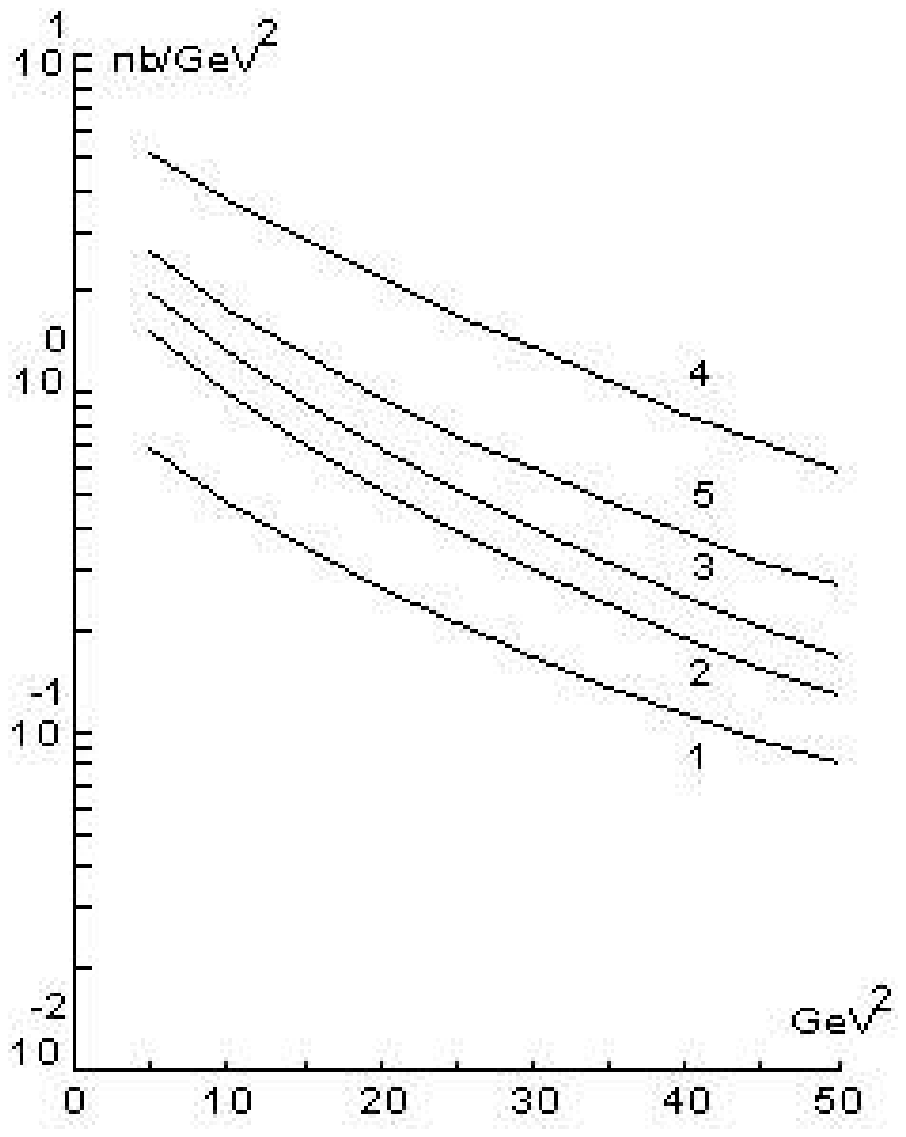,width=10cm,height=10cm}
\end{center}
\centerline{Fig.~9.}
\label{fig9}
\end{figure}
\begin{figure}
\begin{center}
\epsfig{figure=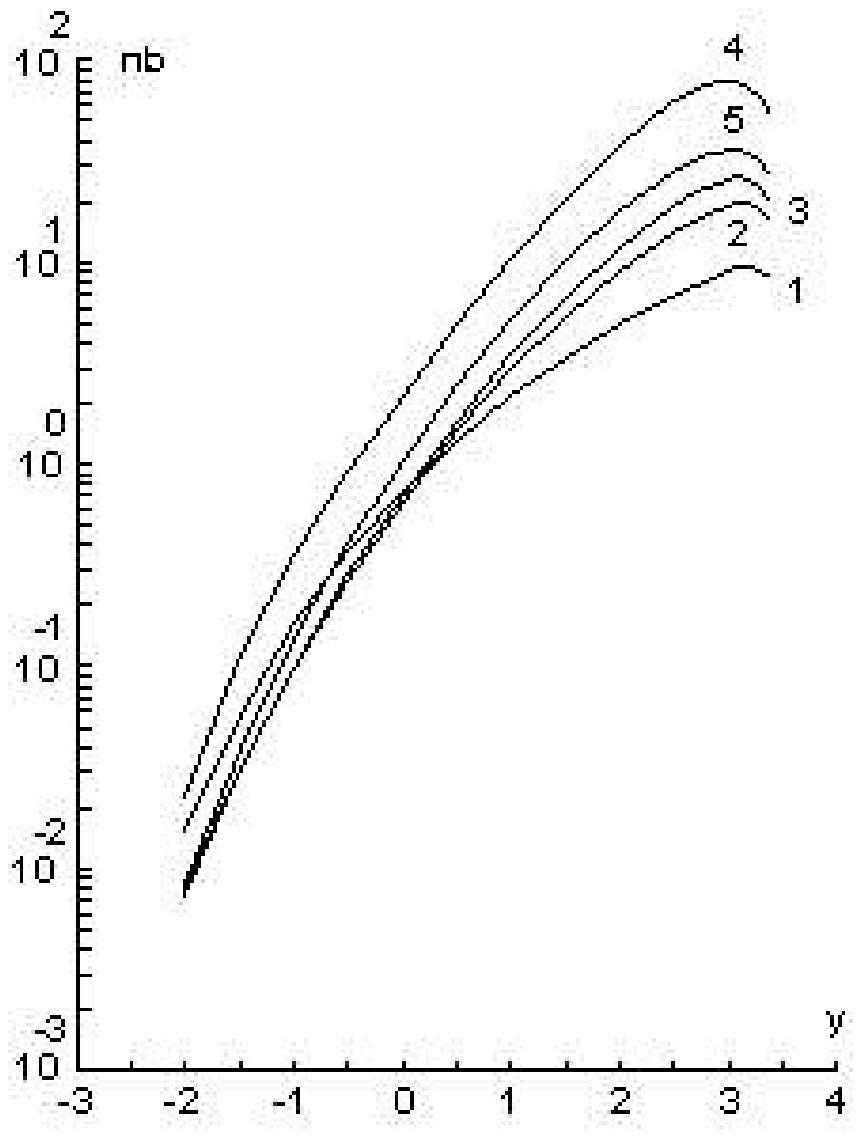,width=10cm,height=10cm}
\end{center}
\centerline{Fig.~10.}
\label{fig10}
\end{figure}

\end{document}